\documentclass[
reprint,           % journal's actual layout
superscriptaddress,% authors with affiliations via superscripts
amsmath,           % add AMS-Latex features
amssymb,           % add extra AMS symbols, including amsfonts
prl,               % prl, pra, prb, prc, prd, pre, prstab
notitlepage,       % control appearance of title page
longbibliography,  % show article titles in the bibliography
floatfix,          % process floats as early as possible
]{revtex4-1}

\usepackage{tensor}     % manipulate tensors
\usepackage{graphicx}   % include figures
\usepackage[
colorlinks=true,        % color link
citecolor=blue,         % cite color
linkcolor=blue,         % link color
urlcolor=blue           % url color
]{hyperref}             % create hyperlinks
\usepackage{bm}         % \bm{<text>} Bold math symbols
\usepackage{xcolor}     % textcolor
\usepackage{float} %???????????
\usepackage{subfigure} %?????????????
\newcommand{\newc}{\newcommand*} 
\newc{\figurewidth}{3.2in}
\newc{\xbar}{\bar{x}}
\newc{\rhoeq}{\rho_{\rm{eq}}}
\newc{\zeq}{z_{\rm{eq}}}
\newc{\la}{\lambda}
\newc{\tla}{\tilde{\la}}
\newc{\dt}{\delta}
\newc{\Dt}{\Delta}
\newc{\vj}{\mathbf{j}}
\newc{\vl}{\bm{l}}
\newc{\hx}{\hat{x}}
\newc{\hy}{\hat{y}}
\newc{\bj}{\bm{j}}
\newc{\mJ}{\mathcal{J}}
\newc{\mP}{\mathcal{P}}
\newc{\ga}{\gamma}
\newc{\Msun}{M_\odot}
\newc{\app}{\approx}
\newc{\av}[1]{\langle #1 \rangle}
\newc{\eq}[1]{Eq.~\eqref{#1}}
\newc{\al}{\alpha}
\newc{\Xstar}{X_{\ast}}
\newc{\seq}{\sigma_{\rm{eq}}}
\newc{\fpbh}{f_{\rm{pbh}}}
\newc{\VT}{\langle VT \rangle}

\def\({\left(}
\def\){\right)}
\def\[{\left[}
\def\]{\right]}

\def\e{\begin{equation}}
\def\q{\end{equation}}
\def\m{\begin{eqnarray}}
\def\n{\end{eqnarray}}

\begin{document}

\title{Secular evolution of compact binaries revolving around a spinning massive black hole}

%%%%%%%%%%%%%%%%%%%%%%%%%%%%%%% author 1 %%%%%%%%%%%%%%%%%%%%%%%%%%%%%
\author{Yun Fang}
\email{fangyun@itp.ac.cn}
\affiliation{CAS Key Laboratory of Theoretical Physics, 
Institute of Theoretical Physics, Chinese Academy of Sciences,
Beijing 100190, China}
\affiliation{School of Physical Sciences, 
University of Chinese Academy of Sciences, 
No. 19A Yuquan Road, Beijing 100049, China}

%%%%%%%%%%%%%%%%%%%%%%%%%%%%%%% author 2 %%%%%%%%%%%%%%%%%%%%%%%%%%%%%
\author{Qing-Guo Huang}
\email{huangqg@itp.ac.cn}
\affiliation{CAS Key Laboratory of Theoretical Physics, 
Institute of Theoretical Physics, Chinese Academy of Sciences,
Beijing 100190, China}
\affiliation{School of Physical Sciences, 
University of Chinese Academy of Sciences, 
No. 19A Yuquan Road, Beijing 100049, China}
%\affiliation{Center for Gravitation and Cosmology, College of Physical Science and Technology, Yangzhou University, Yangzhou 225009, China}
%\affiliation{Synergetic Innovation Center for Quantum Effects and Applications, Hunan Normal University, Changsha 410081, China}

\date{\today}
%%%%%%%%%%%%%%%%%%%%%%%%%%%%%%%%%%%%%%%%%%%%%%%%%%%%%%%%%%%%%%%%%%%%%%
\begin{abstract}

The leading order effect of the spin of a distant supermassive black hole (SMBH) on the orbit of compact binary revolving around it appears at the 1.5 post Newtonian (PN) expansion through coupling with the inner and outer orbital angular momenta, and makes an oscillating characteristic contribution to the secular evolution of the compact binary eccentricity over a long time scale compared to the Kozai-Lidov dynamics caused by a non-spinning SMBH.  

%In this letter we work out the leading order influence of the spin of faraway SMBH on the orbit of the compact binary around it. The spin's affect firstly appear at the 1.5 PN in the dynamics which is from the coupling of the spin angular momentum with the orbital angular momentum of inner and outer orbit. This couple will result in a non-conservation of the total orbital angular momentum of this system, as a result, the relative angle of the inner and outer orbital longitude of ascending note will enter into the dynamics. The numerical results of the secular evolution of the inner orbital eccentricity combined the Newtonian quadrupole results with the spin's correction indicate that spin will cause a change both in the amplitude and phase of kozai-Lidov effect, where the change of phase will accumulate with the increase of time.    

\end{abstract}

\pacs{???}

\maketitle

%%%%%%%%%%%%%%%%%%%%%%%%%%%%%%%%%%%%%%%%%%%%%%%%%%%%%%%%%%%%%%%%%%%%%%

The first direct detection of gravitational wave (GW) by the Laser Interferometric Gravitational-wave Observatory (LIGO) \cite{Abbott:2016blz} is a historical landmark. It provides a new way to explore the Universe, and opens the era of gravitational wave astronomy. 

Up to now, several GW events produced by the inspiral and subsequent merger of two black holes (BHs) or two neutron stars were reported in \cite{Abbott:2016nmj, Abbott:2017vtc, Abbott:2017gyy, Abbott:2017oio, TheLIGOScientific:2017qsa, TheLIGOScientific:2016pea, LIGOScientific:2018mvr}. The signal of GW150914 was strong enough to be apparent, without using any waveform model, in the filtered detector strain data. But the other GW signals were so weak that the matched filtering with waveform templates was essential for the detection. The templates adopted by LIGO in searching for GW signals assumed the circular obits for the compact binaries \cite{Abbott:2017vtc, Abbott:2017gyy, Abbott:2017oio, TheLIGOScientific:2017qsa, TheLIGOScientific:2016pea, LIGOScientific:2018mvr}. This assumption is reasonable for the isolated binary systems because the angular momentum that gravitational waves carry away causes the orbits to circularize faster than they shrink \cite{Peters:1963ux}. 

The real environment where binary BHs live in are often complicated and they are mostly living in multiple systems. Hierarchical triple system is the simplest and stable one, where binary BHs might obtain large eccentricity perturbed by the third body through Kozai-Lidov mechanism \cite{Kozai:1962zz, Lidov:1962zs}. The formation channel of binary BHs roughly fall into two categories: they are from the remanent of evolvement of isolated stellar binaries in ``field" chanel \cite{Voss:2003ep}, or they are formed by ``dynamically" channels through three body encounters in dense star clusters \cite{Sigurdsson:1993}. Triple system of stars are believed to be common in universe \cite{ Raghavan:2010ds, Tokovinin:2014ds, Fuhrmann:2017df, Tokovinin:2006jm, Pribulla:2006gk}, it is possible that triple BHs form in the end of the evolution of triple star systems. On the other hand, the dynamical formation requires a very high stellar density which are believed to exist within the cores of globular clusters (GCs), and the BHs being more massive than the average stars sink to the center of GC until the the majority of the BHs reside in the cluster core \cite{Spitzer:1969df}. This ``mass segregation" process guaranteed that three body encounters in the core can frequently occur \cite{Ivanova:2005mi}, producing binary BHs at high rates \cite{Rodriguez:2015oxa}.

Kozai-Lidov mechanism has an important secular effect in hierarchical triple systems, and plays an important role in dynamical evolution of triples. GWs emitted by highly eccentric orbits of compact binaries excited from the Kozai-Lidov mechanism might be detectable by LIGO and VIRGO \cite{Wen:2002km, OLeary:2005vqo, Mandel:2007hi}, pulsar timing arrays \cite{Finn:2010ph, AmaroSeoane:2009af, Kocsis:2011ch}, and also future space-based GW observatories such as LISA \cite{Robson:2018svj}. Even though Kozai-Lidov mechanism has been extended to more general cases \cite{Katz:2011hn, Naoz:2012bx, Naoz:2010xi, Will:2017vjc}  in the past decades, the effect of spin is absent in the literature. Actually the black holes in our Universe universally have spins. In particular,  the observations mainly based on X-ray Reflection Spectroscopy \cite{Reynolds:2013qqa, Reynolds:2013rva} indicate that most of SMBHs especially with masses larger than $10^7 M_\odot $ have large spins (e.g. the dimensionless spin parameter larger than 0.9). 

In this letter, we will explore the secular evolution of a compact BH binary revolving around a spinning SMBH. At the leading order the effect of spin of SMBH can be interpreted by the gravitomagnetic force \cite{Nichols:2011pu} which induces couplings between spin of SMBH and the inner and outer orbital angular momenta, and leads to an oscillating characteristic correction to the secular evolution of the compact binary eccentricity over a long time scale compared to the Kozai-Lidov dynamics caused by a non-spinning SMBH.
%Then the total orbital angular momentum is not conserved and the relative angle of the inner and outer orbital longitude of ascending node becomes dynamical. Our results indicate that both the amplitude and phase of the secular evolution of the compact binary eccentricity will change due to the spin of SMBH compared to Kozai-Lidov mechanism without spin, and this change caused by spin  will oscillate periodically in a large time scale.

\begin{figure}[H] %H??????!htb????????htbp?????
\centering %????
\includegraphics[height=6.3cm]{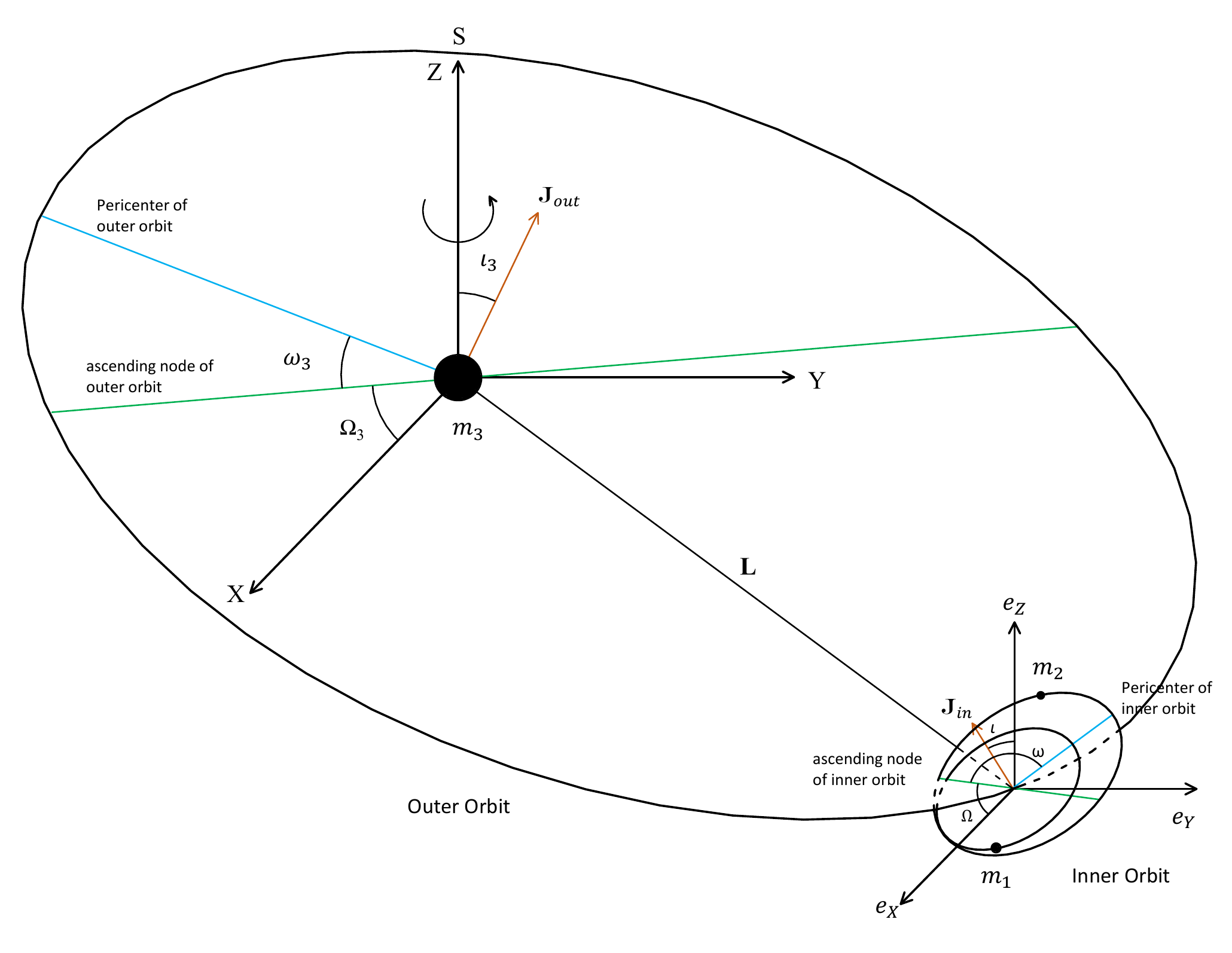}  %?????[]???????????????
\caption{Hierarchical triple system orbits viewed in the fundamental reference frame} %??????????????
\label{Figure.triple} %?????????
\end{figure}

Let's start with a hierarchical triple system illustrated in Fig.~\ref{Figure.triple}, where two binary BHs with mass $m_1$ and $m_2$ revolve around a spinning SMBH with mass $m_3 (\gg m_1, m_2)$ and spin $a$. 
Here the spin of SMBH is taken to align the $Z$ direction in the fundamental reference frame whose origin is located at the position of SMBH. ${\bf{S}} =m_3 a {\bf{e}}_Z$ denotes the spin vector of SMBH, and ${\bf{J}}_{\text{in}}$ and ${\bf{J}}_{\text{out}}$ are the orbital angular momentum of the inner and outer orbits respectively. The green lines are the lines of ascending nodes,  and the blue lines are the lines of pericenters. The coordinates of these two small BHs are denoted by $\mathbf{r}_{\beta}$, $\mathbf{r}_{\beta\gamma}\equiv \mathbf{r}_{\beta}-\mathbf{r}_{\gamma}$, and $\mathbf{n}_{\beta\gamma}={\mathbf{r}_{\beta\gamma}/ r_{\beta\gamma}}$ with $r_{\beta\gamma}\equiv |\mathbf{r}_{\beta\gamma}|$,  where $\beta, \gamma=(1, 2)$. The coordinate $\bf{L}$ of the mass center of the inner binary satisfies $m_1 \mathbf{r}_1+m_2 \mathbf{r}_2=m \bf{L}$, where $m=m_1+m_2$. Introducing $\bm{\ell}=\mathbf{r}_{12}$, we have $\mathbf{r}_1={m_2\over m}\bm{\ell}+\mathbf{L}$, $\mathbf{r}_2=-{m_1\over m}\bm{\ell}+\mathbf{L}$, and then the velocities of these two small black holes read  
\e
{\mathbf{v}_1}={d\mathbf{r}_1\over dt}={m_2\over m}\mathbf{v}+\mathbf{V},\  {\mathbf{v}_2}={d\mathbf{r}_2\over dt}=-{m_1\over m}\mathbf{v}+\mathbf{V}, 
\label{limitv}
\q
where $\mathbf{V}={d\bm{L}/ dt}$, $\mathbf{v}={d\bm{\ell}/ dt}$. 

The interactions induced by spin in binary or multiple systems have been well investigated in post Newtonian (PN) expansion in the weak field and low velocity limit in \cite{Thorne:1984mz, Barker:1975ae, DEath:1975wqz, Barker:1979, Thorne:1980ru, Poisson:1997ha}. At the leading order the spin of SMBH makes a contribution to the equation of motion of the compact binary BHs in 1.5 PN which can be explained as the gravitomagnetic force \cite{Nichols:2011pu} from the space-time of spinning SMBH. The accelerations of $m_1$ and $m_2$ BHs due to the gravitomagnetic force caused by the spin of SMBH take the form
\e
\mathbf{a}^{[1.5\text{PN,spin}]}_{\beta} =2 a m_3\mathbf{v}_{\beta}\times {(\mathbf{e}_Z-3{(\mathbf{e}_Z \cdot \mathbf{r}_{\beta})\mathbf{r}_{\beta}\over r^2_{\beta}})\over r^3_{\beta}}, 
\label{gravitoforce}
\q 
where $a\equiv J_3/m_3$ and $J_3$ is the angular momentum of black hole $m_3$. 
In order to figure out the leading order contribution to the dynamics of inner binary from spin of SMBH, we decompose the equations of motion of inner and outer orbits up to 1.5 PN as follows 
\m
\mathbf{a}&=&\mathbf{a}^{[\text{N}]}+\mathbf{a}^{[1\text{PN}]}+\mathbf{a}^{[1.5\text{PN,without spin}]}+\mathbf{a}^{[1.5\text{PN,spin}]}, 
\label{accel}\\
\mathbf{A}&=&\mathbf{A}^{[\text{N}]}+\mathbf{A}^{[1\text{PN}]}+\mathbf{A}^{[1.5\text{PN, without spin}]}+\mathbf{A}^{[1.5\text{PN, spin}]}, 
\label{Accel}
\n
where $\mathbf{a}\equiv {d^2\bm{\ell}/d t^2}$, $\mathbf{A}\equiv{d^2\bm{L}/ d t^2}$. Here $\mathbf{a}^{[\text{N}]}$ and $\mathbf{A}^{[\text{N}]}$ are the Newtonian accelerations for the inner and outer orbits, namely 
\m
\mathbf{a}^{[N]}&=&-m{ \mathbf{n}\over {\ell}^2}-m_3{\mathbf{r}_1 \over {|\mathbf{r}_1|}^3}+m_3{\mathbf{r}_2\over {|\mathbf{r}_2|}^3},
\label{innermotion} \\
\mathbf{A}^{[N]}&=&-{m_1\over m}{m_3\mathbf{r}_{1}\over {|\mathbf{r}_1|}^3}-{m_2\over m}{m_3\mathbf{r}_{2}\over {|\mathbf{r}_2|}^3}. 
\label{outermotion}
\n
Up to the quadrupole order, we have 
\m
{\bf{a}}^{[\text{N}]}&=&-{m \over \ell^2}{\bf{n}}+{m_3 \ell \over L^3}(3 N_n {\bf{N}}-{\bf{n}}), 
\label{innermotionquad}\\
{\bf{A}}^{[\text{N}]}&=& -{m_3 \over L^2} {\bf{N}} - {3\over 2}{m_3 {\ell}^2\over L^4}{m_1 m_2\over m^2}((5{N_n}^2-1) {\bf{N}}-2N_n {\bf{n}} ), 
\label{outermotionquad}
\n
where $\bf{n}={\bm{\ell}/ \ell}$, ${\bf{N}}={{\bf{L}}/ L}$, and $N_n=\bf{N}\cdot\bf{n}$. 
From Eq.~(\ref{gravitoforce}), the leading contributions to the accelerations of the relative motion and center of mass of inner binary BHs from the spin of SMBH are given by 
\m 
\mathbf{a}^{[1.5\text{PN,spin}]}&\simeq&2 a m_3\mathbf{v}\times {(\mathbf{e}_Z-3{(\mathbf{e}_Z \cdot \mathbf{N})\mathbf{N}})\over L^3},
\label{inneracceleration}\\
 \mathbf{A}^{[1.5\text{PN,spin}]}&\simeq&2 a m_3\mathbf{V}\times {(\mathbf{e}_Z-3{(\mathbf{e}_Z \cdot \mathbf{N})\mathbf{N}})\over L^3}, 
\label{outeracceleration}
\n 
where the higher order corrections of ${\cal O}({\ell/ L})$ are ignored.

The inner and outer orbits are dominated by the Keplerian orbits, and the acceleration in (\ref{accel}) and (\ref{Accel}) are dominated by $a^{[\text{N}]}$ and $A^{[\text{N}]}$ respectively. The inner orbital elements $p, e, \omega, \Omega, \iota $ are defined as 
\m
\ell&=&{p\over 1+e\cos{\phi}},\ \bm{\ell}=\ell \mathbf{n},\ \bm{\lambda}={d\mathbf{n}\over d\phi},\ \hat{\mathbf{h}}=\mathbf{n}\times\bm{\lambda}, \\
\mathbf{n}&=&[\cos{\Omega}\cos{(\omega+\phi)}-\cos{\iota}\sin{\Omega}\sin{(\omega+\phi)}]\mathbf{e}_X \nonumber \\
&+&[\sin{\Omega}\cos{(\omega+\phi)}+\cos{\iota}\cos{\Omega}\sin{(\omega+\phi)}]\mathbf{e}_Y \nonumber \\
&+&\sin{\iota}\sin{(\omega+\phi)}\mathbf{e}_Z, 
\n
and the semi-major axes of inner orbit is $\alpha=p(1-e^2)$.
Similarly, for the outer orbit, we have 
\m
L&=&{P\over 1+E\cos{\Phi}},\ \mathbf{L}=L \mathbf{N},\ \bm{\Lambda}={d\mathbf{N}\over d\Phi},\ \hat{\mathbf{H}}=\mathbf{N}\times\bm{\Lambda}, \\
\mathbf{N}&=&[\cos{\Omega_3}\cos{(\omega_3+\Phi)}-\cos{\iota_3}\sin{\Omega_3}\sin{(\omega_3+\Phi)}]\mathbf{e}_X \nonumber\\
&+&[\sin{\Omega_3}\cos{(\omega_3+\Phi)}+\cos{\iota_3}\cos{\Omega_3}\sin{(\omega_3+\Phi)}]\mathbf{e}_Y \nonumber\\
&+&\sin{\iota_3}\sin{(\omega_3+\Phi)}\mathbf{e}_Z, 
\n 
and the semi-major axes of outer orbit is $A=P(1-E^2)$. 

According to Eqs.~(\ref{accel}), (\ref{Accel}), (\ref{innermotionquad}) and (\ref{outermotionquad}), we introduce the perturbing accelerations $\delta \mathbf{a}=\mathbf{a}+{m\over \ell^2}\mathbf{n}$ and $\delta \mathbf{A}=\mathbf{A}+{m_3\over L^2}\mathbf{N}$. The dynamical evolution of the inner orbital elements are govern by the following equations of motion 
\e
{d\mathbf{h}\over dt}=\bm{\ell}\times {\delta \mathbf{a}}, m{d\mathbf{R}\over dt}=\delta \mathbf{a} \times \mathbf{h}+\mathbf{v}\times (\bm{\ell}\times \delta \mathbf{a}), 
\q
where $\mathbf{h}\equiv \bm{\ell}\times \mathbf{v}=\sqrt{m p}\hat{\mathbf{h}}$, $\mathbf{R}$ is the Runge-Lenz vector defined by $\mathbf{R}\equiv {\mathbf{v}\times \mathbf{h}/ m}-\mathbf{n}=e (\cos{\phi} \mathbf{n}-\sin{\phi}\bm{\lambda})$. Then the evolution of the orbital elements can be obtained by solving the above two equations, namely 
\e
\begin{split}
{dp\over dt}=&2\sqrt{p^3\over m}{\mathcal{S}\over 1+e \cos{\phi}}, \\
{de\over dt}=&\sqrt{p\over m}(\sin{\phi}~\mathcal{R}+{2\cos{\phi}+e+e{\cos{\phi}}^2\over 1+e \cos{\phi}}\mathcal{S}), \\
{d\bar{\omega}\over dt}=&{1\over e}\sqrt{p\over m}(-\cos{\phi}~\mathcal{R}+{2+e\cos{\phi}\over 1+e \cos{\phi}}\sin{\phi}~\mathcal{S}), \\
{d\iota\over dt}=&\sqrt{p\over m}{\cos{(\omega+\phi)}\over 1+e \cos{\phi}}\mathcal{W}, \\
\sin{\iota}{d\Omega\over dt}=&\sqrt{p\over m}{\sin{(\omega+\phi)}\over 1+e \cos{\phi}}\mathcal{W}, 
\end{split}
\label{innerelements}
\q
where $\mathcal{R}=\mathbf{n}\cdot\delta\mathbf{a}$, $\mathcal{S}=\bm{\lambda}\cdot \delta\mathbf{a}$, $\mathcal{W}=\hat{\mathbf{h}}\cdot \delta\mathbf{a}$, $\bar{\omega}$ is defined by $\dot{\omega}=\dot{\bar{\omega}}-\dot{\Omega}\cos{\iota}$. 
Similarly, the outer orbital elements evolution can be expressed by replacing all the elements of inner binary by the the elements of outer binary, like $e \to E, p \to P, m \to m_3,\phi \to \Phi$, $\dot{\omega}_3=\dot{\bar{{\omega}}}_3-\dot{\Omega}_3\cos{\iota_3}$, etc., and $\mathcal{R}_3=\mathbf{N}\cdot\delta\mathbf{A}$, $\mathcal{S}_3=\mathbf{\Lambda}\cdot \delta\mathbf{A}$, $\mathcal{W}_3=\hat{\mathbf{H}}\cdot \delta\mathbf{A}$. 

The secular evolution of orbital elements are double averaged results through integrating over one orbital period of both inner and outer orbits as follows 
\e
\langle F \rangle= {1\over T_{\text{out}}}{1\over T_{\text{in}}}\int_{0}^{T_{\text{out}}}\int_{0}^{T_{\text{in}}}Fdt dt', 
\label{averaget}
\q
where $T_{\text{in}}$ and $T_{\text{out}}$ are the periods of inner and outer orbits. For convenience, the time integration can be replaced by phase angle integration according to the relation of $dt=\sqrt{p^3/ m}(1+e\cos{\phi})^{-2}d\phi$ and $dt'=\sqrt{P^3/ m_3}(1+E\cos{\Phi})^{-2}d\Phi$, and thus the average in Eq.~(\ref{averaget}) becomes 
\e
\begin{split}
\langle F \rangle=& {1 \over 4 \pi^2} (1-e^2)^{3/2} (1-E^2)^{3/2}\\
&\int_{0}^{2\pi}\int_{0}^{2\pi}{ F \over (1+e\cos{\phi})^{2} (1+E\cos{\Phi})^{2} }d\phi d\Phi, 
\end{split}
\q
For simplicity, we also convert the time derivation ${d/ dt}$ to a dimensionless one ${d/d\tau}$ by rescaling time compared to the inner orbital period with $\tau\equiv t/T_{\text{{in}}}={t \over 2\pi}\sqrt{m\over {\alpha}^3}$. 

\begin{figure}[H] %H??????!htb????????htbp?????
\centering %????
\includegraphics[height=5.3cm]{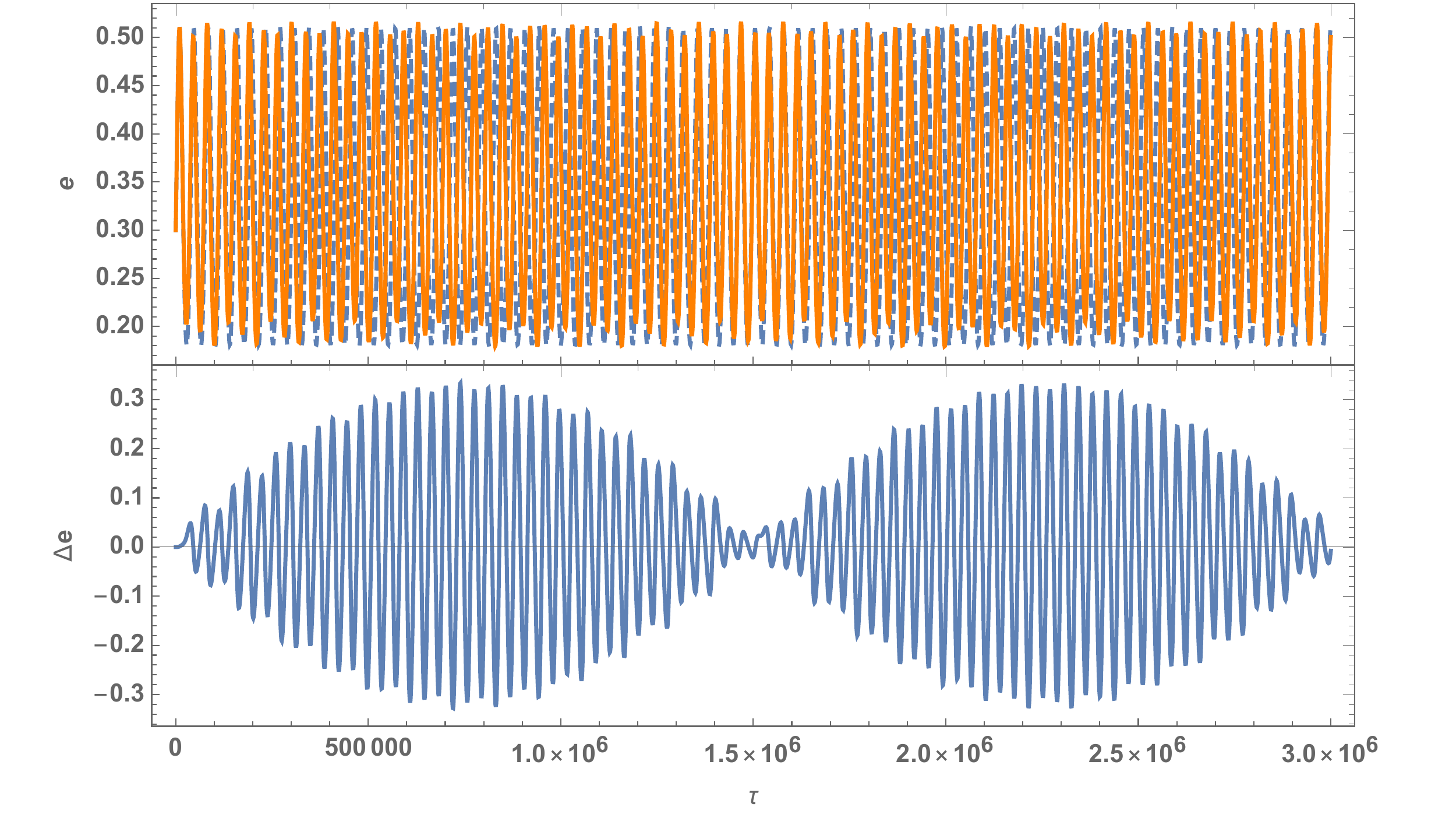}
\caption{The evolution of the inner orbital eccentricity $e$, where $\Delta e$ encodes the difference between the total effects including the spin of SMBH given in this letter (the orange curve in the upper panel) and the Kozaio-Lidov dynamics (the blue dashed curve in the upper panel). Here the initial data are given by $m_1=m_2=10 M_\odot, m_3=10^7 M_\odot, a=0.9m_3, \alpha=0.03 AU, A=100AU, e=0.3, E=0.5, \omega={\pi\over 3}, \iota={\pi\over 3}, \iota_3={\pi\over 12}, \Omega={\pi\over 32}, \Omega_3={\pi\over 6}$.  }%??????????????
\label{Figure.etau} %?????????
\end{figure}

The Newtonian quadrupole perturbing accelerations in Eqs.~(\ref{innermotionquad}) and (\ref{outermotionquad}) result in the well-known Kozai-Lidov effect as follows
\begin{widetext}
    \begin{eqnarray}
\begin{split}
{de \over d\tau}=&{15 \pi  \alpha ^3 e \sqrt{1-e^2} m_3 \over 16 A^3 (1-{E}^2)^{3/2}(m_1+m_2)}(\sin ^2\iota _3 (\cos 2 \iota+3) \sin 2 \omega \cos (2 \Omega -2 \Omega_3)+4 \sin ^2\iota _3 \cos \iota  \cos 2 \omega \sin (2 \Omega -2 \Omega _3)\\
&-4 \sin 2 \iota _3 \sin \iota  \cos 2 \omega \sin (\Omega -\Omega _3)-2 \sin 2 \iota \sin 2 \iota _3 \sin 2 \omega \cos (\Omega -\Omega _3)+\sin ^2\iota  (3 \cos 2 \iota _3+1) \sin 2 \omega), \\
{d \iota\over d \tau}=& {3 \pi  \alpha ^3 m_3 \over {4 A^3\sqrt{1-e^2} (1-{E}^2)^{3/2} (m_1+m_2)}}(\sin \iota  \sin \iota _3 \cos (\Omega -\Omega _3)+\cos \iota \cos\iota _3) (\sin\iota _3 \sin (\Omega -\Omega _3) (5 e^2 \cos 2 \omega+3 e^2+2)\\
&+5 e^2 \sin 2 \omega (\sin \iota _3 \cos \iota  \cos (\Omega -\Omega _3)-\sin \iota  \cos\iota _3)), \\
{d\Omega\over d\tau}=& {3 \pi  \alpha ^3 m_3 \over {4 A^3 \sqrt{1-e^2}(1-{E}^2)^{3/2} (m_1+m_2)}}(\sin \iota _3 \cos (\Omega -\Omega _3)+\cos \iota _3 \cot \iota) (5e^2 \sin \iota _3 \sin 2 \omega \sin (\Omega -\Omega _3)\\
&+(5 e^2 \cos 2 \omega-3 e^2-2) (\sin \iota \cos \iota _3-\sin \iota _3 \cos \iota\cos (\Omega -\Omega _3))), \\
{d \bar{\omega}\over d\tau}=&{3 \pi  \alpha ^3 \sqrt{1-e^2} m_3 \over {8 A^3 (1-{E}^2)^{3/2} (m_1+m_2)}}(10 \sin \iota \sin 2 \iota _3 \sin 2 \omega \sin (\Omega -\Omega _3)\\
&+\sin^2\iota _3 \cos (2\Omega -2\Omega _3) (2 \sin ^2\iota (4-5 \cos^2 \omega)+20 \cos^2 \omega-10)-10 \sin^2 \iota _3 \cos \iota \sin 2 \omega \sin (2\Omega -2\Omega _3)\\
&+\sin 2 \iota \sin 2 \iota _3 (3-5 \cos 2 \omega) \cos (\Omega -\Omega _3)+(3 \cos 2 \iota _3+1) (\sin ^2\iota(5 \cos^2 \omega-4)+1)), \\
{d E\over d\tau}=&0, \\
{d\iota_3\over d\tau}=&-{3 \pi  \alpha ^{7/2} m_1 m_2 \sqrt{m_3} \over {4 A^{7/2} (1-{E}^2)^2(m_1+m_2)^{5/2}}}(\sin \iota _3 (\sin (2 \Omega -2 \Omega _3) (\sin ^2\iota  (-5 e^2 \cos ^2\omega+4 e^2+1)+10 e^2 \cos ^2\omega-5 e^2)\\
&+5 e^2 \cos \iota  \sin 2 \omega  \cos (2 \Omega -2 \Omega_3))+\cos \iota _3 (\sin 2 \iota \sin (\Omega -\Omega _3) (-5 e^2 \cos ^2\omega+4 e^2+1)-5e^2 \sin \iota  \sin 2 \omega  \cos (\Omega -\Omega _3))), \\
{d{{\Omega}}_3\over d\tau}=& -{3 \pi  \alpha ^{7/2} m_1 m_2 \sqrt{m_3} \csc\iota _3 \over {8 A^{7/2} (1-{E}^2)^2 (m_1+m_2)^{5/2}}}(\sin 2 \iota _3 (\frac{1}{2} \sin ^2\iota (\cos ( 2\Omega -2 \Omega _3)+3) (5 e^2 \cos 2 \omega -3 e^2-2)\\
&+5 e^2 \cos \iota  \sin 2 \omega  \sin( 2\Omega -2 \Omega _3)-5 e^2 \cos 2 \omega  \cos ( 2\Omega -2 \Omega _3)+3 e^2+2)\\
&+\cos 2\iota _3 (\sin 2 \iota \cos (\Omega -\Omega _3) (5 e^2 \cos 2 \omega -3 e^2-2)-10 e^2 \sin \iota  \sin 2\omega \sin (\Omega -\Omega _3))), \\
{d{\bar{\omega}}_3\over d\tau}=&{3 \pi  \alpha ^{7/2} {m_1} {m_2} \sqrt{{m_3}} \over {16 A^{7/2} (1-{E}^2)^2 ({m_1}+{m_2})^{5/2}}}(30 e^2 \sin \iota\sin 2\iota _3 \sin 2\omega \sin (\Omega -\Omega _3)-30 e^2 \sin ^2\iota _3 \cos \iota \sin 2\omega \sin (2 \Omega -2 \Omega _3)\\
&+3 \sin ^2\iota _3 \cos (2 \Omega -2 \Omega _3) (\sin ^2\iota (-5 e^2 \cos 2\omega+3 e^2+2)+10 e^2 \cos 2\omega)\\
&+3 \sin 2\iota \sin 2\iota _3 \cos (\Omega-\Omega _3) (-5 e^2 \cos 2\omega+3 e^2+2)+(2-3 \sin ^2\iota _3) (\sin ^2\iota (15 e^2 \cos 2\omega-9e^2-6)+6 e^2+4))\\
%&{3 \pi  \alpha ^{7/2} m_1 m_2 \sqrt{m_3} \over {128 A^{7/2} (1-{E}^2)^2 (m_1+m_2){}^{5/2}}} (240 e^2 \sin \iota  \sin 2 \iota _3 \sin 2 \omega \sin (\Omega -\Omega_3)-240 e^2 \sin ^2\iota _3 \cos \iota  \sin 2 \omega \sin (2\Omega -2\Omega _3)\\
%&+6 \sin^2\iota _3 \cos (2\Omega -2\Omega _3) (10 e^2 (\cos 2 \iota+3) \cos 2 \omega+4 (3 e^2+2) \sin ^2\iota )\\
%&+24 \sin 2 \iota \sin 2 \iota _3 \cos (\Omega -\Omega _3) (-5 e^2 \cos 2\omega+3 e^2+2)\\
%&+(3 \cos 2 \iota _3+1) (60 e^2 \sin ^2\iota  \cos 2 \omega+2 (3 e^2+2) (3 \cos 2 \iota+1))), \\
\end{split}
 \label{elementsgeneral}
    \end{eqnarray}
  \end{widetext}
Here both the inner orbital angular momentum $\mathbf{J}_{\text{in}}={m_1 m_2\over m}\sqrt{pm}\hat{\mathbf{h}}$ and outer orbital angular momentum $\mathbf{J}_{\text{out}}={m}\sqrt{Pm_3}\hat{\mathbf{H}}$ can be in arbitrary directions, and the above results come back to the familiar Kozai-Lidov formular when the total orbital angular momentum $\mathbf{J}=\mathbf{J}_{\text{in}}+\mathbf{J}_{\text{out}}$ is along the $Z$ direction. 
%where $\Omega_3$, $\Omega$ will cancel out by the relation of $\Omega_3=\Omega+\pi$. The canceling of $\Omega_3$, $\Omega$ in the dynamics is a result of the conservation of the total orbital angular momentum. 
%The higher order orbital motion of hierarchical three body system studied before \cite{Katz:2011hn, Naoz:2012bx, Naoz:2010xi, Will:2017vjc} have not contained spin. Without spin, the total orbital angular momentum is conserved and the longitude of ascending nodes do not enter into the dynamics. 
According to Eqs.~(\ref{inneracceleration}) and (\ref{outeracceleration}), the spin of SMBH $\mathbf{S}$ couples with $\mathbf{J}_{\text{in}}$ and $\mathbf{J}_{\text{out}}$ in 1.5 PN at the leading order. 
%In the presence of spin of the centering SMBH in our case, from $\mathbf{J}_{in}={m_1 m_2\over m}\bm{\ell}\times \mathbf{v}$ and $\mathbf{J}_{out}=m\mathbf{L}\times \mathbf{V}$, Eq.~(\ref{inneracceleration}) and (\ref{outeracceleration}) tells there is a couple of $\mathbf{J}_{in}$ and $\mathbf{J}_{out}$ with $\mathbf{S}$ at the leading 1.5 PN order. The change of $\mathbf{J}_{in}$ and $\mathbf{J}_{out}$ can be obtained from the relation $\mathbf{J}_{in}={m_1 m_2\over m}\mathbf{h}$ and $\mathbf{J}_{out}={m}\mathbf{H}$ with $\mathbf{H}=\sqrt{m_3 P}\hat{\mathbf{H}}$, then ${d\mathbf{J}_{in}\over dt}={m_1 m_2\over m}\bm{\ell}\times \mathbf{\delta \mathbf{a}}$ and ${d\mathbf{J}_{out}\over dt}={m}\mathbf{L}\times \mathbf{\delta \mathbf{A}}$. With the perturbing force in Eq.~(\ref{inneracceleration}) and (\ref{outeracceleration}), we see that ${d\mathbf{J}\over dt}\neq0$, thus the total orbital angular momentum is not conserved any more when coupled with spin.   
%In principle, the spin angular momentum will also change at 1.5 PN motion, but because spin angular momentum is far more larger than the total orbital angular momentum, the change of spin angular momentum can be neglected. $Z$ axis of the reference frame could be set in the direction of spin for simplicity. 
The perturbing forces in Eqs.~(\ref{inneracceleration}) and (\ref{outeracceleration}) from the leading order effect of spin contribute to the double averaged results of secular evolution of orbital elements as follows 
\e
\begin{split}
{de \over d\tau}=&{d E\over d\tau}={d p\over d\tau}={d P\over d\tau}={d \iota_3\over d\tau}\supset 0, \\
{d \iota\over d \tau}\supset&{3 \pi  a \alpha ^{3/2} {m_3} \sin 2 \iota _3 \sin (\Omega -\Omega _3)\over 2A^3 (1-{E}^2)^{3/2} \sqrt{{m_1}+{m_2}}}, \\
{d\Omega\over d \tau}\supset& -{\pi  a \alpha ^{3/2} {m_3}  \over 2 A^3(1-{E}^2)^{3/2} \sqrt{{m_1}+{m_2}}}\\
&\times (-3 \sin2 \iota _3 \cot\iota \cos (\Omega -\Omega _3)+3 \cos2 \iota _3+1), \\
{d\omega\over d\tau}\supset& -{3 \pi  a \alpha ^{3/2} {m_3} \sin2 \iota _3 \csc\iota \cos (\Omega-\Omega _3) \over 2 A^3 (1-{E}^2)^{3/2} \sqrt{{m_1}+{m_2}}}, \\
{d\Omega_3\over d \tau}\supset&{4 \pi  a \alpha ^{3/2} {m_3}\over A^3 (1-{E}^2)^{3/2}\sqrt{m_1+m_2}}, \\
{d \omega_3\over d\tau}\supset&-{12 \pi  a \alpha ^{3/2} {m_3} \cos\iota _3 \over A^3 (1-{E}^2)^{3/2}\sqrt{{m_1}+{m_2}}}, 
\label{leadingspin}
\end{split}
\q
where ``$\supset$" indicates the contribution from the spin only. It apparently seems that the spin of SMBH does not lead to the secular evolution of the eccentricity of the compact binary. However, Eqs.~(\ref{leadingspin}) couples to the those in Eqs.~(\ref{elementsgeneral}), and the spin of SMBH influence the eccentricity of the compact binary through the effects on the other orbital elements, such as $\iota$, $\omega$, $\Omega-\Omega_3$ and etc., which are closely related to the eccentricity $e$ of the inner orbit in Eqs.~(\ref{elementsgeneral}). 
For a numerical example, see Fig.~\ref{Figure.etau}.   
Fig.~\ref{Figure.etau} indicates that the spin of SMBH leads to a correction to the evolution of inner orbital eccentricity in both the amplitude and phase. Furthermore, such a correction oscillates over a long time scale.

In this letter we sketch out the leading order contribution to the secular evolution of the compact binary revolving around a spinning SMBH. The leading order interactions induced by the spin of SMBH appear at the 1.5 PN through coupling between spin and the inner and outer orbital angular momenta. The relative angle between the inner orbital longitude of ascending node and the outer orbital longitude of ascending node becomes dynamical. Through affecting the evolutions of some orbital elements, the spin of SMBH finally influence 
the secular evolution of the orbit eccentricity of compact binary. Compared to the Kozai-Lidov mechanism caused by a non-spinning SMBH, the spin of SMBH makes an oscillating characteristic contribution to the secular evolution of the compact binary eccentricity over a long time scale. 

Finally, the features on the secular evolution of the orbit elements of compact binary from the spin of SMBH leaves some fingerprints in the GW waveform which might be used to measure the spin of SMBH in the future.

Acknowledgments. 
This work is supported by grants from NSFC 
(grant No. 11690021, 11575271, 11747601), 
the Strategic Priority Research Program of Chinese Academy of Sciences 
(Grant No. XDB23000000), Top-Notch Young Talents Program of China, 
and Key Research Program of Frontier Sciences of CAS. 
%%%%%%%%%%%%%%%%%%%%%%%%%%%%%%%%%%%%%%%%
%%%%%%%%%%%%%%%%%%%%%%%%%%%%%%%%%%%%%%%%

%%%%%%%%%%%%%%%%%%%%%%%%%%%%%%%%%%%%%%%%
%%%%%%%%%%%%%%%%%%%%%%%%%%%%%%%%%%%%%%%%

%%%%%%%%%%%%%%%%%%%%%%%%%%%%%%%%%%%%%%%%
%%%%%%%%%%%%%%%%%%%%%%%%%%%%%%%%%%%%%%%%
\end{document}